\documentclass[a4paper,11pt]{article}
\usepackage{pos}

\def\svev#1{\left\langle #1\right\rangle}       

\newcommand{\bee}{\begin{equation}}
\newcommand{\ee}{\end{equation}}
\newcommand{\beea}{\begin{eqnarray}}
\newcommand{\eea}{\end{eqnarray}}

\title{Approaching the Chiral and Continuum Limit of Large-N QCD}

\author*[a]{Evan Wickenden}
\author[a]{Thomas DeGrand}

\affiliation[a]{
Department of Physics,
University of Colorado, \\ Boulder, CO 80309, USA}

\emailAdd{evan.wickenden@colorado.edu}
\emailAdd{thomas.degrand@colorado.edu}

\abstract{
	We present preliminary results 
	from our calculation of the low energy constants  (LECs) of the chiral effective theory for
	3, 4 and 5 color QCD with
	$N_f=2$ dynamical fermion flavors. We simulate with clover fermions over a range of lattice couplings and
	quark masses. 
	We observe the expected $N_c$ scaling
	for the LECs  appropriate to the condensate $B$ and pseudoscalar decay constant  $F$.
	The range of quark masses over which leading order chiral perturbation theory describes the data grows as $N_c$ rises.
}

\FullConference{%
 The 38th International Symposium on Lattice Field Theory, LATTICE2021
  26th-30th July, 2021
  Zoom/Gather@Massachusetts Institute of Technology
}

\begin{document}
\maketitle

\newcommand{\comment}[1]{{\color{red}#1}}

\section{Introduction}

\newcommand{\SU}{\text{SU}}

The large $N_c$ limit provides interesting insight into nonperturbative features of QCD
\cite{tHooft:1973alw, tHooft:1974pnl,Witten:1980sp}.  The lattice offers a means to test its
predictions.  The objective of our study is to simulate   across $N_c$  with $N_f = 2$ flavors of
dynamical fermions, measure the low energy constants (LECs) of chiral perturbation theory
($\chi$PT), and observe their $N_c$ scaling.  We aim to extrapolate our results  to $a \to 0$ and
$N_c \to \infty$.

The results shown here are preliminary.   We observe the expected $N_c$ scaling for the leading
LECs: $B \sim N_c^0$ and $F \sim N_c^{1/2}$. A feature we observe is that the region of parameter
space where $\chi$PT describes the data is larger at larger $N_c$. This is due to the fact that the chiral expansion may
be written in terms of the parameter $\xi = \frac {m_\pi^2}{8\pi^2 f_\pi^2}$, which, because
$f_\pi^2 \sim N_c$, is naturally suppressed at large $N_c$.

A large $N_c$ study with $N_f = 2$ flavors in the chiral and continuum limit is absent from the
literature. A number of other approaches to the large $N_c$ limit have been studied.
Hernandez et al. considered the $N_f=4$ case with $3-6$ colors at a single lattice
spacing \protect{\cite{Hernandez:2019qed}}. Bali et al. studied spectroscopy  with 2-7 and 17 colors in 
the quenched approximation \cite{Bali:2013kia}. Using 
Eguchi-Kawai reduction, according to which finite-volume effects vanish as $N_c \to \infty$,
Gonzalez-Arroyo and collaborators have performed studies at extremely large $N_c$
\cite{Perez:2020vbn}. Our goal is to make comparisons with these and other studies.

\section {Wilson Chiral Perturbation Theory}

$\chi$PT has a set of LECs which characterize the quark mass dependence of quantities such as
$m_\pi^2$, $f_\pi$. Predictions for these observables through NNLO are present in the literature. They are
typically presented using
two different expansions: the $x$  and $\xi$ expansions, where 
\begin{align}
	x \equiv \frac {2 B m_q} {4 \pi^2 F^2}, && \xi \equiv \frac {m_\pi^2} {4 \pi f_\pi^2}
\end{align}
Using $\xi$ rather than $x$ amounts to a resummation of the chiral expansion. Careful studies in the
$\SU(3)$ case with $N_f = 2 (+1)$ fermions demonstrate that the $\xi$ expansion has better
convergence properties, with the NNLO formulae describing lattice data to larger fermion or pseudoscalar masses 
\cite{FlavourLatticeAveragingGroup:2019iem,
JLQCD:2008zxm,
Budapest-Marseille-Wuppertal:2013vij}.
A theoretical justification  for this behavior is that the NNLO chiral logs in the $\xi$ expansion have smaller numerical
coefficients than they do in the $x$ expansion. Following these results, we work primarily with the
$\xi$ expansion. 
We plan to do fits with the $x$ expansion as a consistency check. 

QCD on the lattice has chiral symmetry explicitly broken by both nonzero quark mass and nonzero
lattice spacing.  Wilson Chiral Perturbation Theory (W$\chi$PT) enhances chiral perturbation theory
at the level of the effective Lagrangian by introducing operators with an explicit dependence on
lattice spacing. These operators arise from the Symanzik action for QCD via a spurion analysis.
Using the power counting scheme $\delta \sim p^2 \sim m_q \sim a$, Bar, Rupak and Shoresh derive the
effective chiral Lagrangian for Wilson-type fermions through $\mathcal O( \delta^2 )$ 
\cite{Bar:2003mh}.
The NLO W$\chi$PT predictions for $m_\pi^2/m_q$ and $f_\pi$ are:
\begin{align}
	\frac {m_\pi^2} {m_q} &= 2 B (1 + \frac  12 \xi \log \xi) + L_M \xi + W_q a  +
	\tilde W_{qq} \frac {a^2} \xi 
	\nonumber\\
	f_\pi &= F(1 - \xi \log \xi) + L_f \xi +  W_F a .
	\label{eq:chformula}
\end{align}
The $W$'s are the correction terms due to the use of Wilson fermions.


We use the Wilson flow parameter $t_0$ to set the scale. $t_0$ depends on the bare parameters in the simulation,
and thus on measured observables. 
The dependence of the gradient flow scale $t_0$
on $m_{PS}^2$ was described by Bar and Golterman \cite{Bar:2013ora}. Using a $t_0$ 
which is mass-dependent (as was done by Ref.~\cite{Ayyar:2017qdf}) does not alter the chiral logarithms, but it
does affect the $L_i$'s in Eq.~\ref{eq:chformula}. We need a mass-independent definition of a lattice spacing,
which we obtain by  interpolating $m_\pi^2 \sqrt{t_0}$ versus $t_0$ to
$0.15$ for our data at a given $\beta$. 

We match scales across $N_c$ in the commonly - used way, taking
\bee
t_0^2 \svev{E(t_0)} = C(N_c)  
\label{eq:flow}
\ee
with
\bee
 C(N_c)= C \left( \frac{3}{8} \frac{N_c^2-1}{N_c}\right),
\label{eq:ce}
\ee
and $C=0.3$  the usual value used in  $SU(3)$.
Our data span a range of $t_0= 1$ to 3 or so, corresponding (with an $SU(3)$ value of $\sqrt{t_0}\sim 0.15$ fm) to
$a=0.09-0.15$ fm.

\section {Simulation and Analysis Methodology}

We use the HMC algorithm with the Wilson gauge action with nHYP links and clover fermions
\cite{Hasenfratz:2001hp}.
We have mostly
simulated on $16^3 \times 32$ lattices, with a few $24^3 \times 32$ volumes used as cross checks. For each data point, we collected
between 400 and several thousand trajectories, keeping every 10th configuration for measurements. A jackknife
analysis is used to obtain errors on and correlations between lattice observables. By increasing the
number of deletions and monitoring how the errors grow, we can roughly estimate how
autocorrelated the data are, and find them to be minimally so. We discard at least the first hundred
trajectories for thermalization, and further test for thermalization by walking out the minimum
trajectory and watching for drifting observables. Most of the data points included in our analysis
have $m_\pi L > 4$; a few are slightly below this threshold. 
The Python package \texttt{gvar} is used to automatically track errors and correlations, alongside \texttt{lsqfit} for
correlated nonlinear least squares fitting \cite{gvar,lsqfit}.

Pion masses, decay constants and  Axial Ward Identity quark masses are all determined in the usual way of fitting
correlators to hyperbolic functions. As a way to reduce systematic errors associated with choosing fit
ranges,  we employ Neil and Jay's Bayesian model-averaging method 
\cite{Jay:2020jkz}.
This weights each fit proportionally
to $e^{-\chi^2/2 + N}$, where $N$ is the number of points included in the fit, which balances a
small chi-squared against a larger fit range.

\section {Chiral Fits}

\begin{figure}
\begin{center}
	\includegraphics[width=0.49\textwidth]{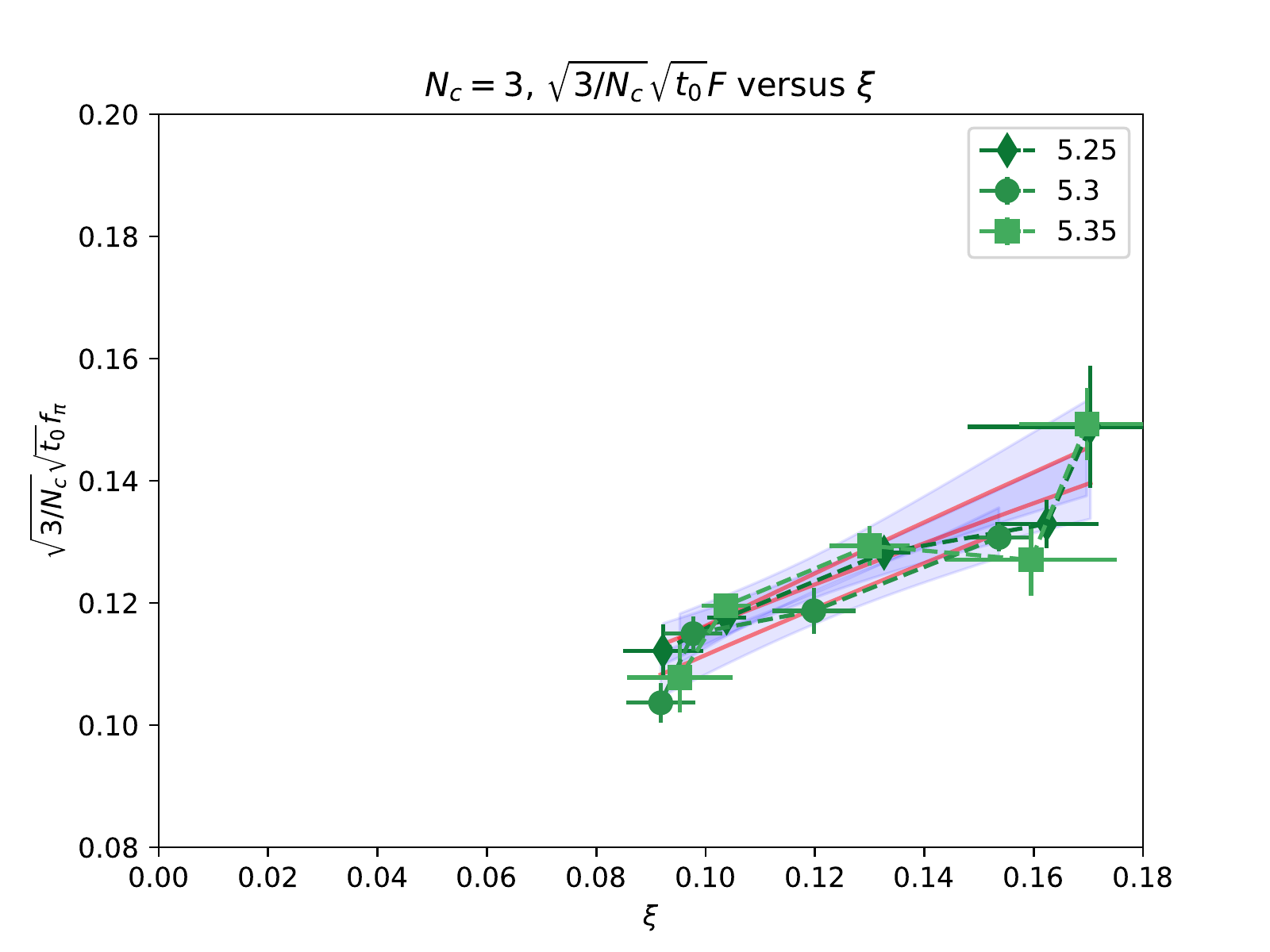}
	\includegraphics[width=0.49\textwidth]{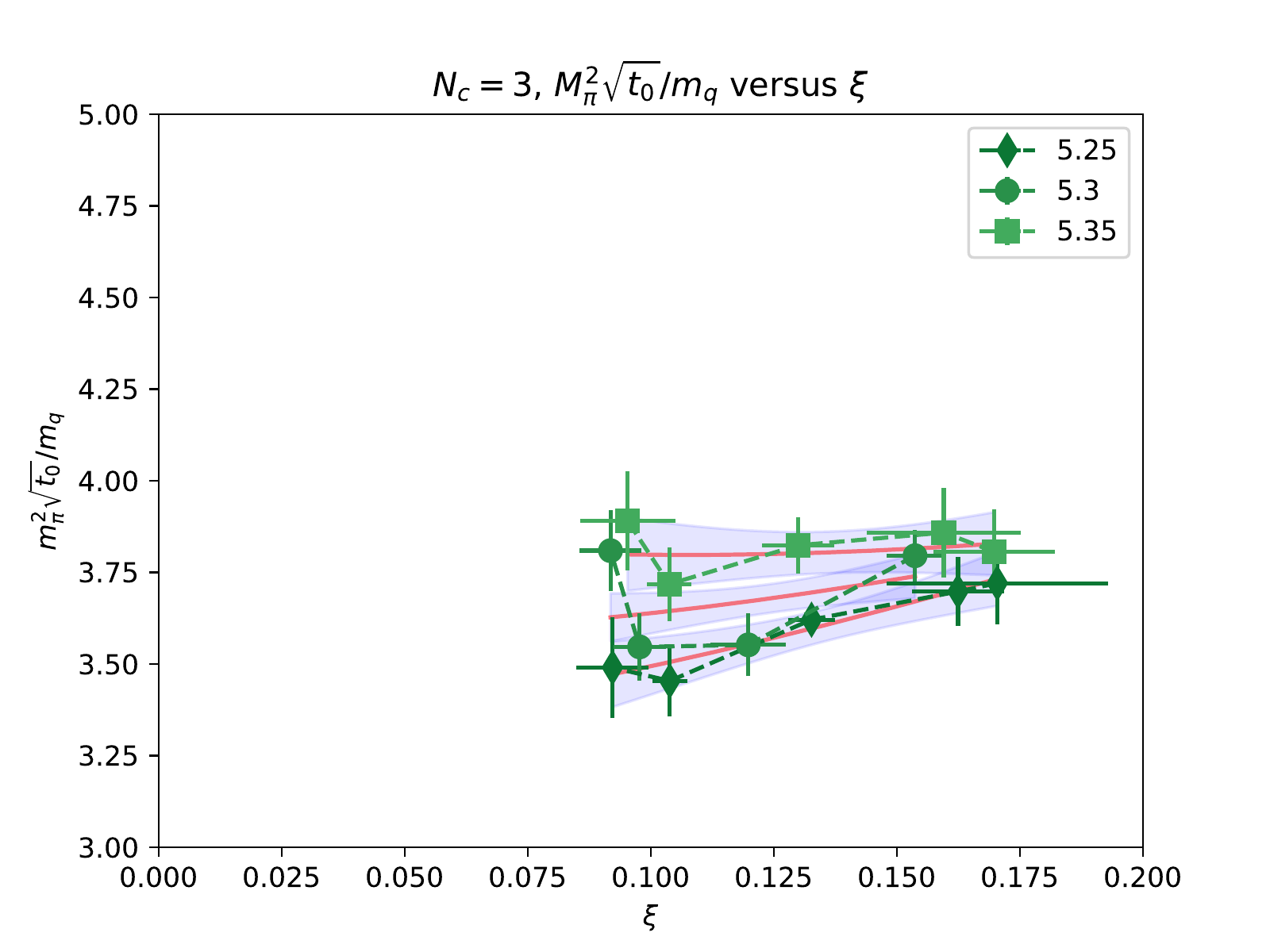}
	\\
	\includegraphics[width=0.49\textwidth]{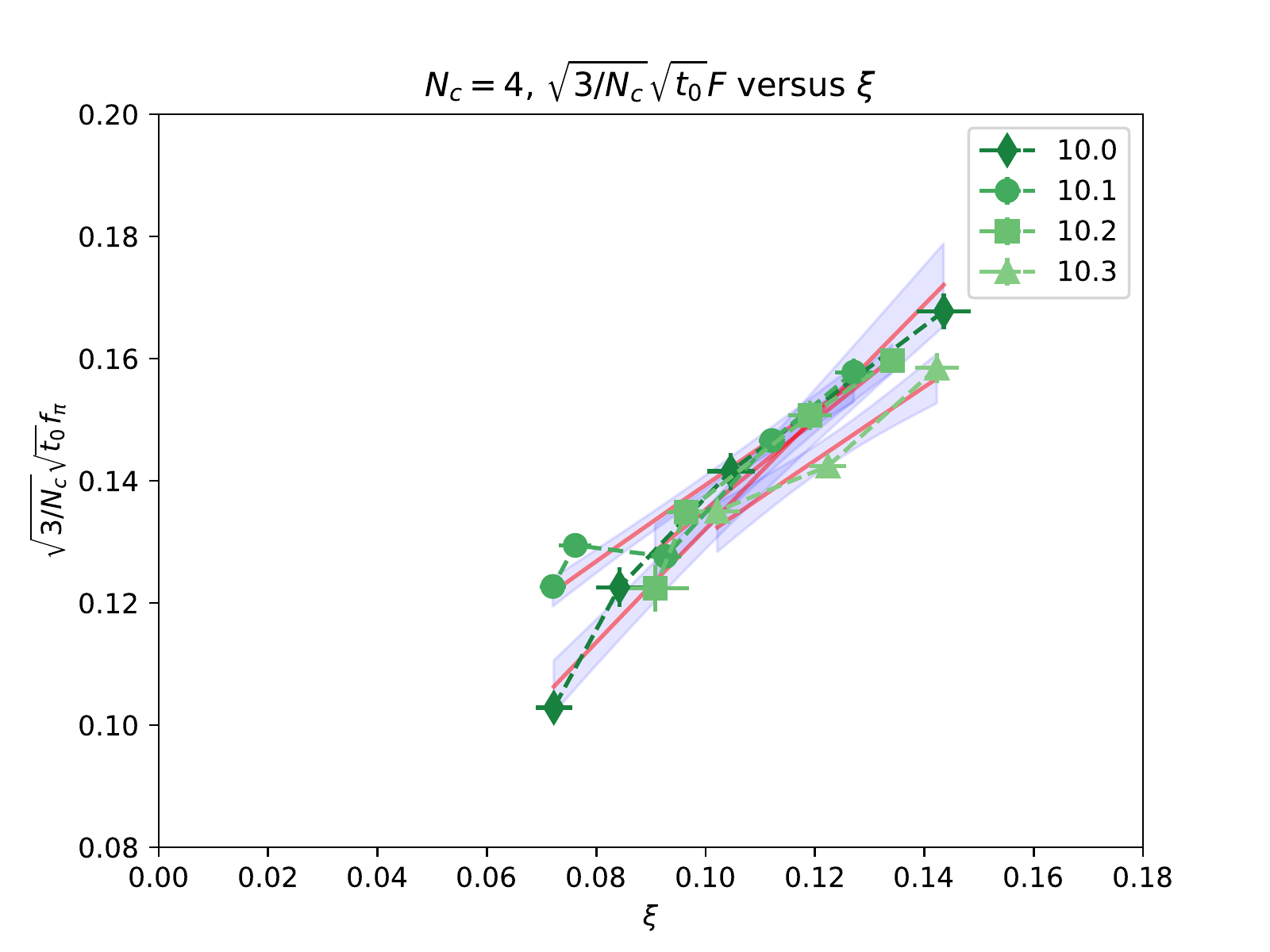}
	\includegraphics[width=0.49\textwidth]{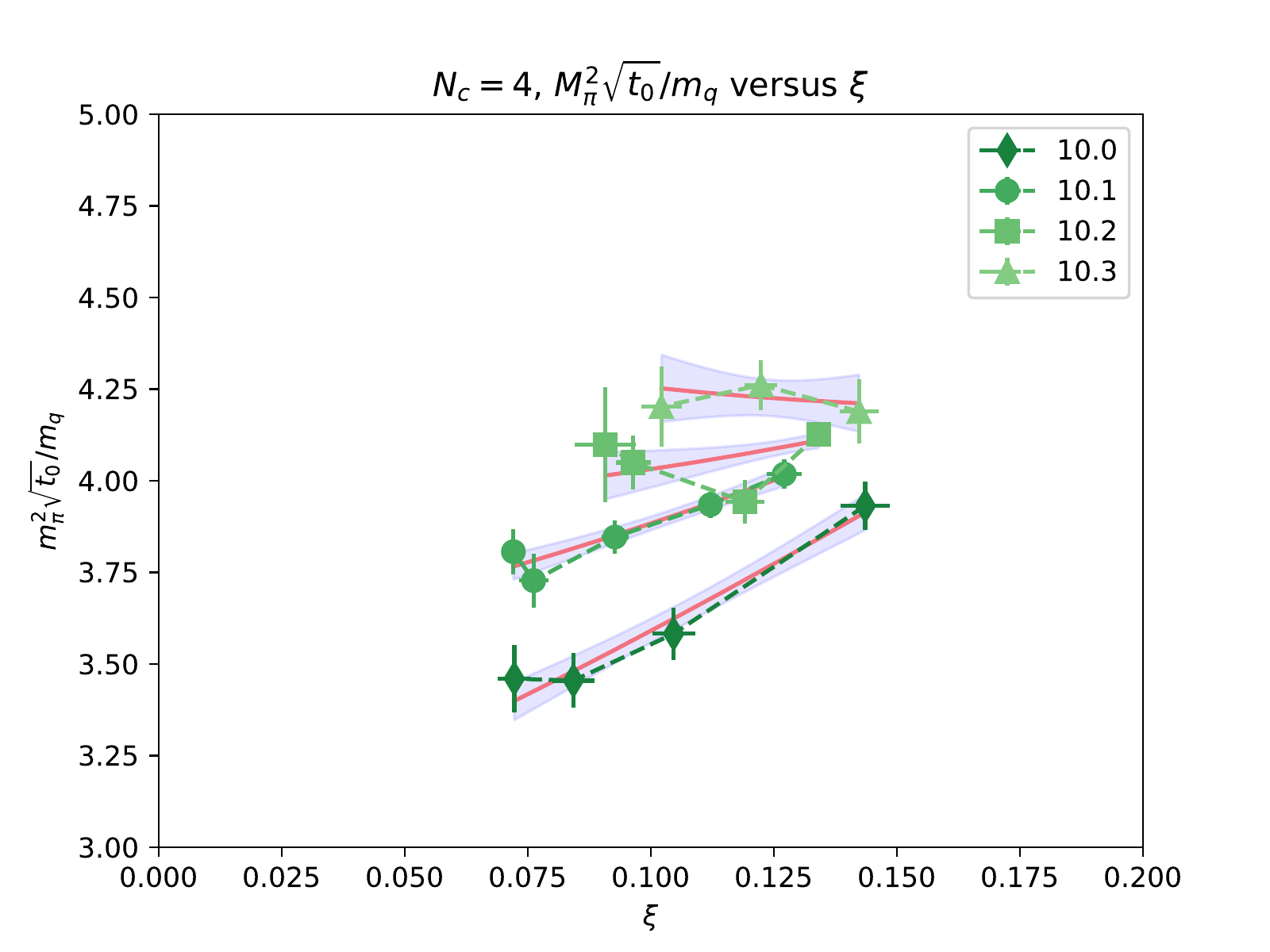}
	\\
	\includegraphics[width=0.49\textwidth]{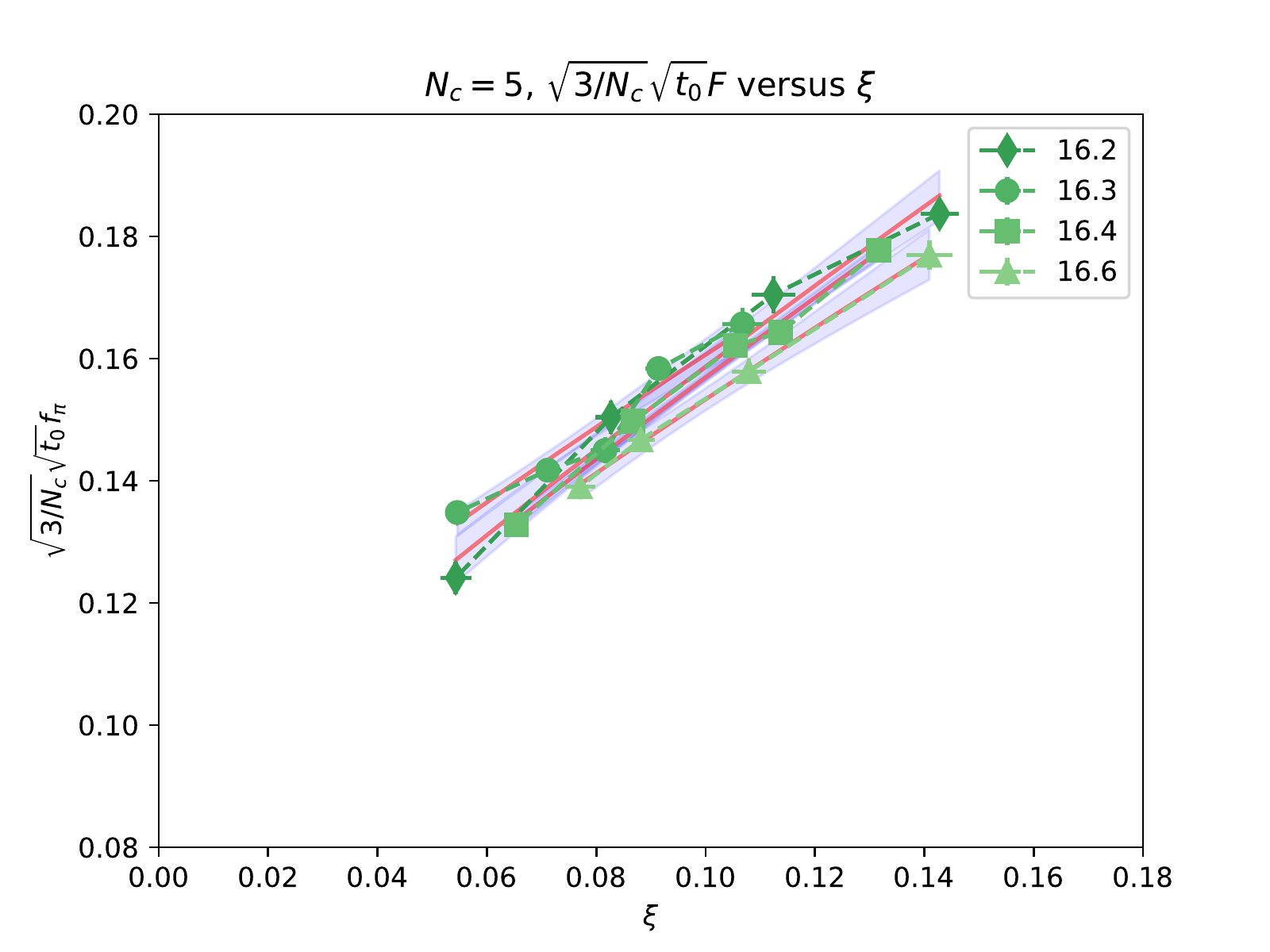}
	\includegraphics[width=0.49\textwidth]{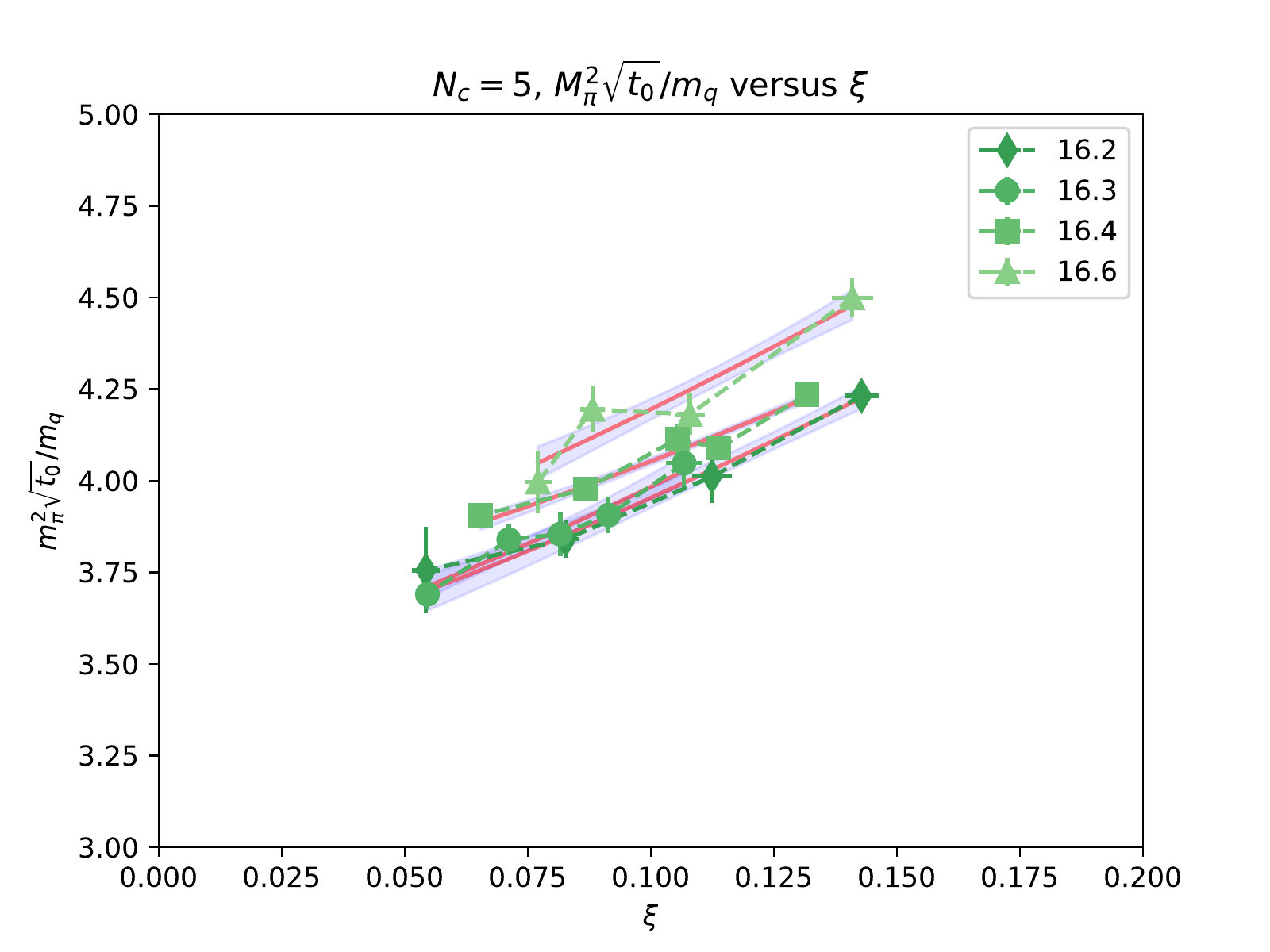}
\end{center}	
	\caption{Left: pion decay constant scaled by  $\sqrt{t_0} \sqrt{3/N_c}$ for 3,4,5 colors. Right: squared
pion mass divided by AWI quark mass (again scaled by $\sqrt{t_0}$).  The lines and colored bands show the results of NLO $\chi$PT fits to the individual
$\beta$ values, neglecting the $W_i$'s in Eq.~\protect{\ref{eq:chformula}}. }
\label{fig:one}
\end{figure}

\begin{figure}
\begin{center}
	\includegraphics[width=0.49\textwidth]{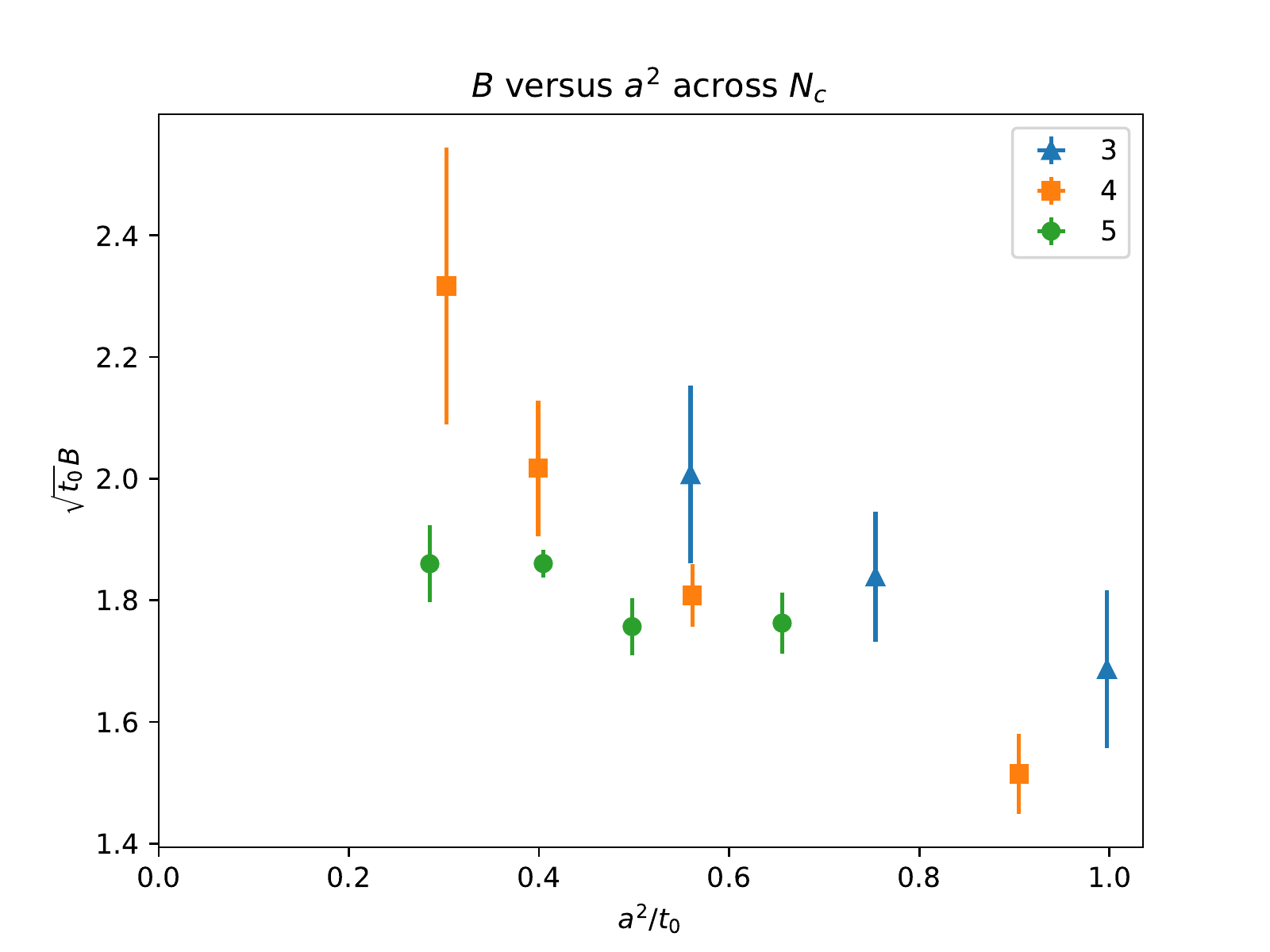}
	\includegraphics[width=0.49\textwidth]{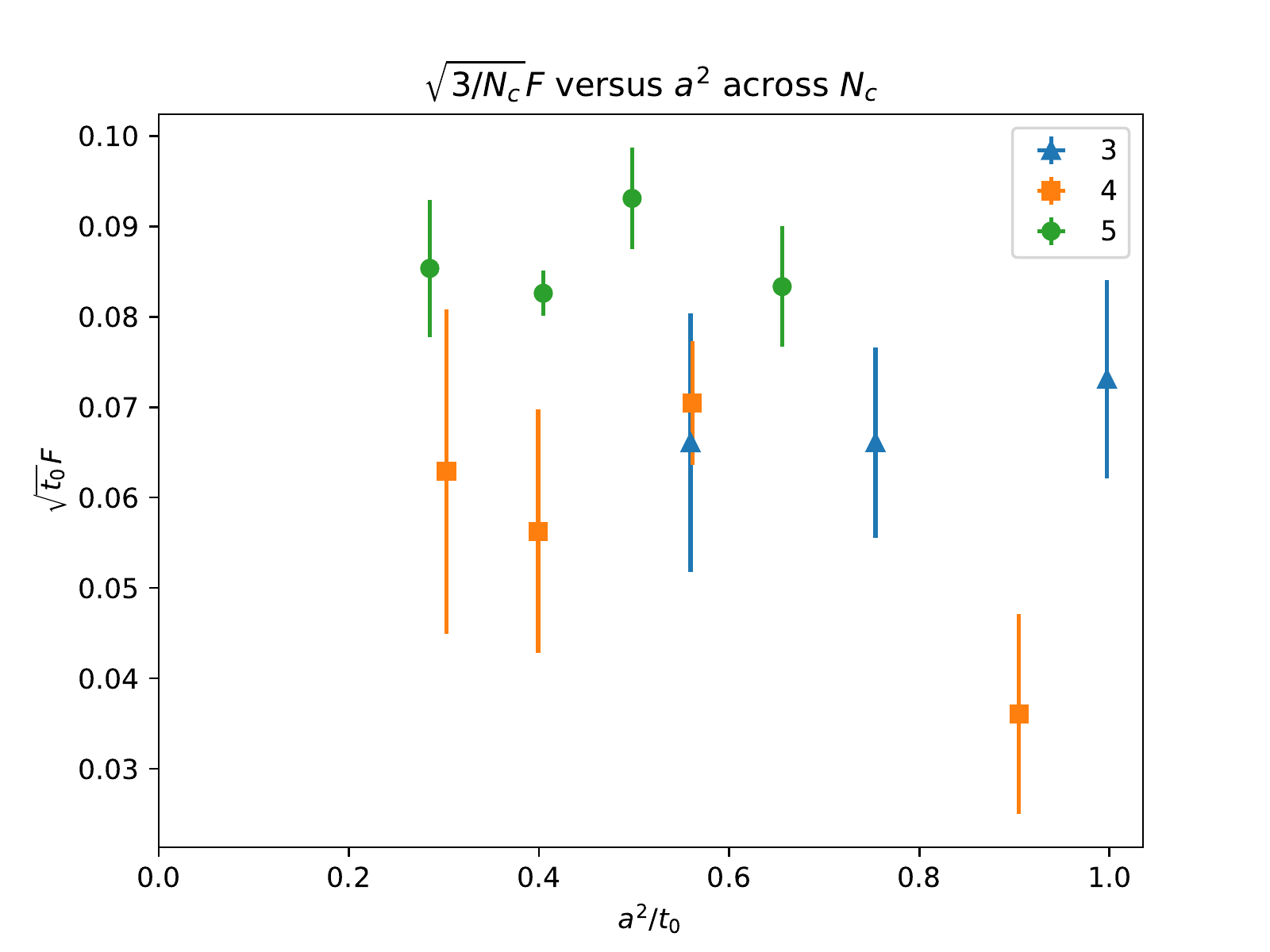}
	\\
	\includegraphics[width=0.49\textwidth]{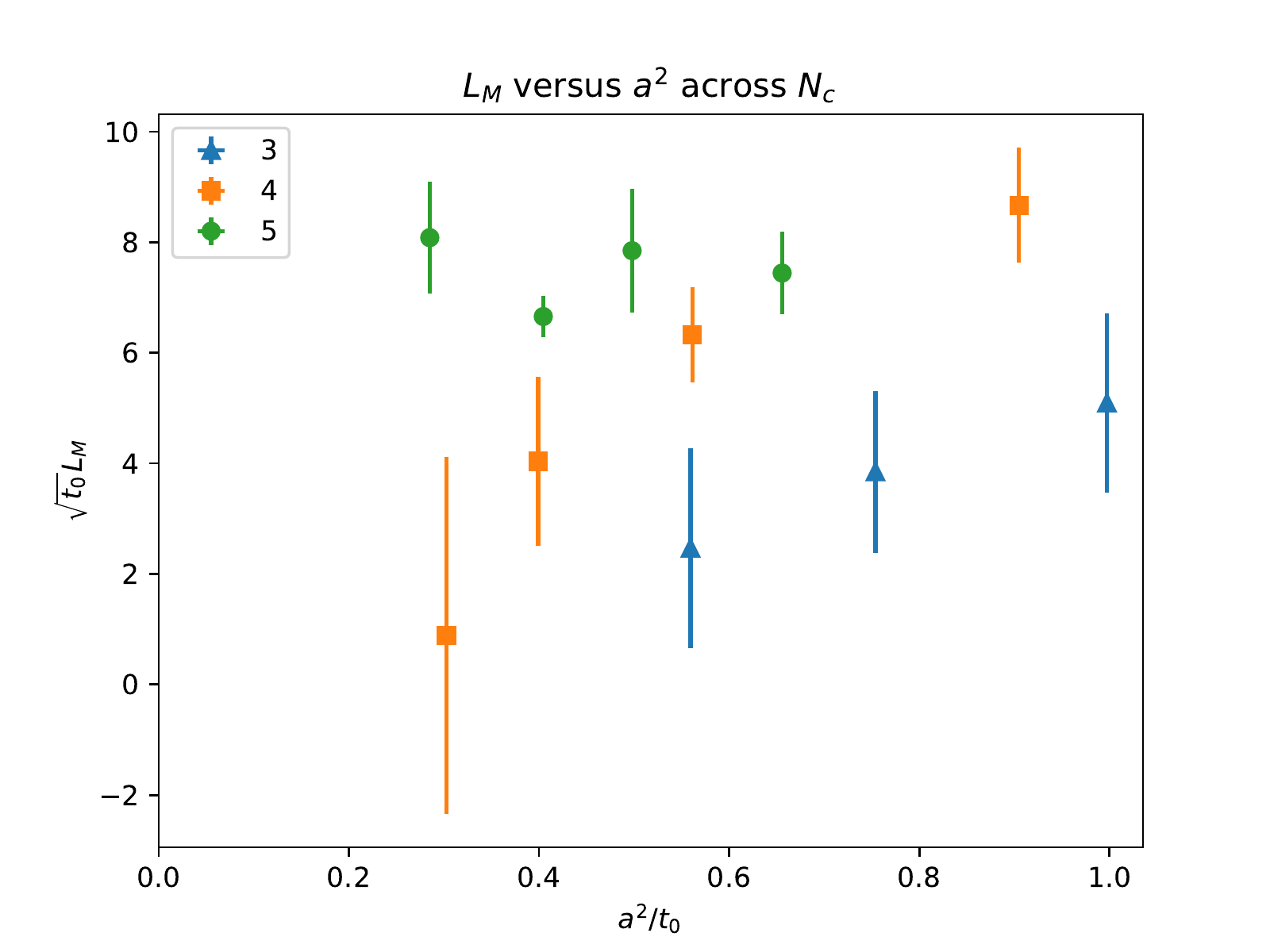}
	\includegraphics[width=0.49\textwidth]{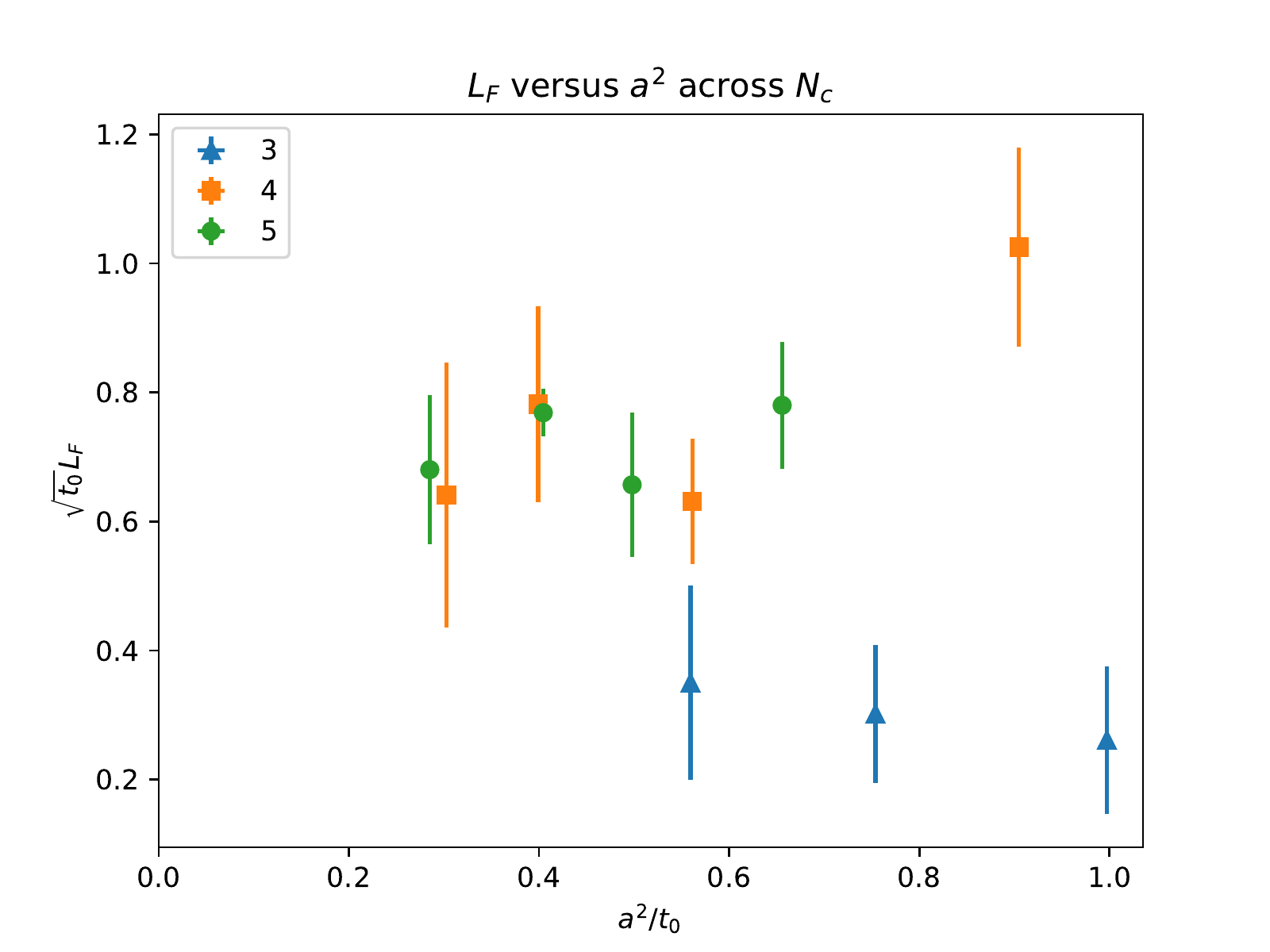}
\end{center}
\caption{Preliminary plots of $B$, $F$, $L_M$, $L_F$  (all in units of $\sqrt{t_0}$) versus $a^2/t_0$. $F$ is scaled by $\sqrt{3/N_c}$.
	Blue points are $\SU(3)$, orange points $\SU(4)$, green points $\SU(5)$. }
\label{fig:two}

\end{figure}

In Fig.~\ref{fig:one} we present plots of $m_\pi^2 \sqrt{t_0} / m_q$ and $f_\pi \sqrt{t_0}$ versus
$\xi$ for 3, 4, and 5 colors. We fit these to standard NLO $\chi$PT at each lattice spacing (i.e.
neglecting the Wilson terms), and in Fig.~\ref{fig:two} plot the $a$-dependence we obtain in $B$,
$F$, $L_M$, and $L_F$. 

\section {Analysis}




The plots for $F$ and $L_F$ show small $a$ dependence and appear to extrapolate to a common value
of $\sqrt{t_0} \sqrt{3/N_c}F \sim 0.07-0.08$ or $\sqrt{3/N_c}F \sim 90-100$ MeV (recall we are working in a convention
where $f_\pi \sim 131$ MeV), or $\sqrt{3/N_c}F \sim 60$- 70 MeV in the ``93 MeV'' $f_\pi$ convention. 
This is a reasonable number, according to the tables in Ref.~\cite{FlavourLatticeAveragingGroup:2019iem}.
$\sqrt{t_0}B$ has larger $a-$ dependence, but a value of $\sqrt{t_0}B\sim 2.0-2.4$ gives $\Sigma^{1/3}\sim 250$ MeV,
again not an unreasonable value.

The above plots also convey a novel feature of this project: the chiral limit becomes
easier at larger $N_c$, in that a constant range of $\xi$ corresponds to a much wider range of
$m_q$. 
Again, this is simply due to the fact that $\xi \sim m_\pi^2 / f_\pi^2 \sim 
m_q / N_c$. This is nevertheless very useful, since it means that at
larger $N_c$, one can obtain good chiral data while staying further away from the
computational challenges of very light dynamical quarks. 
 $\xi$ also governs the size of finite volume effects, so these are also reduced at larger
$N_c$. 
In Fig.~\ref{fig:two}, one can see the LECs are much better determined for $5$ colors than $4$ or $3$ (and
they appear to have less lattice spacing dependence), despite arising from fewer net trajectories.
At three colors, on a $16^3 \times 32$ lattice, we are squeezed between finite volume effects and
keeping $\xi$ small enough for good fits to NLO chiral formulae.  In fact, the bottleneck to our project is
$N_c=3$, which is of course much less interesting than $N_c>3$ due to the availability of much better data by other collaborations.
We only need it to cross check our results.
For this reason we have found it
necessary to begin running on $24^3 \times 32$ lattices.

\section {Conclusion}

In this project we aim to make continuum predictions for the low energy constants of chiral
perturbation theory with $N_f = 2$ dynamical fermions. We aim to assess how well large $N_c$
predictions describe the physical $\text{SU}(3)$ case and to compare our results with other
approaches to the large $N_c$ limit, namely studies with different fermion content---different
$N_f$'s and/or different representations. 

We have so far only used the next-to-leading order prediction of $\text{SU}(N_f)$ Wilson chiral
perturbation theory, with the standard power counting scheme $\delta \sim p^2 \sim m_q \sim a$. In
the large $N_c$ limit, the axial anomaly vanishes and the $\eta$ meson recovers its status as a
Goldstone boson, so $U(N_f)$ $\chi$PT becomes the correct low-energy description of QCD. Considering
alternative power counting schemes, NNLO predicitons, and $U(N_f)$ $\chi$PT are directions for
future study. 

Data collection and analysis continue. The goal remains first $\SU(2) \times \SU(2)$ fits at
	each $N_c$, then global $U(2) \times U(2)$ fits incorporating all data across $N_c$.

\begin{acknowledgments}
This material is based upon work supported by the U.S. Department of Energy, Office of Science, Office of
High Energy Physics under Award Number DE-SC-0010005.
Some of the computations for this work were also carried out with resources provided by the USQCD
Collaboration, which is funded
by the Office of Science of the U.S.\ Department of Energy
using the resources of the Fermi National Accelerator Laboratory (Fermilab), a U.S.
Department of Energy, Office of Science, HEP User Facility. Fermilab is managed by
  Fermi Research Alliance, LLC (FRA), acting under Contract No. DE- AC02-07CH11359.
Our computer code is based on the publicly available package of the
  MILC collaboration~\cite{MILC}. The version we use was originally developed by Y.~Shamir and
  B.~Svetitsky.
\end{acknowledgments}

\end{document}